\documentclass[aps,pra,twocolumn,floatfix]{revtex4-1}

\pdfoutput=1

\usepackage[utf8]{inputenc}
\usepackage{physics,amssymb,amsmath}
\usepackage{bm}
\usepackage{graphicx}
\usepackage[usenames,dvipsnames]{xcolor}
\usepackage[colorlinks,bookmarks=false,citecolor=blue,linkcolor=cyan,urlcolor=blue]{hyperref}
\graphicspath{{./figures/}}

\begin{document}

\title{Collective self-trapping of atoms in a cavity}
\author{A.~Dombi${}^1$, T.~W.~Clark${}^1$, F.~I.~B.~Williams${}^1$,  F.~Jessen${}^2$, J.~Fortágh${}^2$, D.~Nagy${}^1$, A.~Vukics${}^1$, P.~Domokos${}^1$}

\affiliation{${}^1$ Institute for Solid State Physics and Optics, Wigner Research Centre for Physics,  H-1525 Budapest P.O. Box 49, Hungary}

\affiliation{${}^2$ Physikalisches Institut, Eberhard Karls Universit\"at T\"ubingen, D-72076 T\"ubingen, Germany}

\begin{abstract}
We experimentally demonstrate optical dipole trapping of a cloud of cold atoms by means of a dynamically coupled  mode of a high-finesse cavity. We show that the trap requires a collective action of the atoms, {\it i.e.} a single atom would not be trapped under the same laser drive conditions. The atoms pull the frequency of the mode closer to resonance, thereby allowing the necessary light intensity for trapping into the cavity. The back-action of the atoms on the trapping light mode is also manifested by the non-exponential collapse of the trap.
\end{abstract}

\maketitle

\section{Introduction}
\label{sec:intro}

Experimental study of an ensemble of cold atoms placed in a spatially and spectrally selected, discrete, resonant mode of a high finesse optical cavity can yield useful insight into the dynamics of photon-atom interaction. The dynamics of both the atoms and the light field have to be taken into account for understanding such systems, which is a generic feature of cavity quantum electrodynamics (CQED) \cite{Haroche2006Exploring,Ritsch2013Cold}.  The atoms not only move under the forces from the light field but exhibit a strong back action on the state of the field \cite{Gupta2007Cavity,Brennecke2008Cavity}. While the light field is influenced by the averaged interaction with the atoms, the “optical centre of mass”  \cite{Purdy2010Tunable,Schleier-Smith_optomechanical_2011}, individual atoms communicate between themselves through their mutual coupling to the cavity field mode \cite{vaidya_tunable-range_2018} leading to various collective light scattering and synchronisation effects.

A remarkable example is the self-organization of a homogeneous cloud of cold atoms into an optical lattice, where the lattice potential is entirely sustained by the field scattered into the cavity by the periodically ordered atoms \cite{Domokos2002Collective,Black2003Observation,Slama2007Cavityenhanced,arnold_self-organization_2012}. When realized with ultra-cold atoms \cite{Baumann2011Exploring}, this phenomenon led to the observation of quantum phase transitions in open, driven-dissipative systems  \cite{Nagy2011Critical,Dimer2007Proposed} and of exotic dynamical phases of matter \cite{landig_quantum_2016,leonard_supersolid_2017,cosme_dynamical_2018,georges_light-induced_2018,morales_coupling_2018,dogra_dissipation-induced_2019,schuster_supersolid_2020}. While single-atom cQED plays a leading role in quantum information processing \cite{reiserer_cavity-based_2015}, many-atom CQED experiments have enabled the observation of other fundamental effects such as dynamical phase transitions \cite{mivehvar_emergent_2019,muniz_exploring_2020}, quantum critical phenomena \cite{schmidt_dynamical_2014,nagy_nonequilibrium_2015}, photon thermalisation \cite{schutz_dissipation-assisted_2016,wang_photon_2018}, quantum squeezing \cite{schleier-smith_squeezing_2010,cox_deterministic_2016} and quantum feedback \cite{kroeger_continuous_2020}. Additionally, many-atom ensembles in a strongly coupled cavity can also constitute a non-linear medium  at low photon numbers \cite{chang_quantum_2014}. The discreteness of modes of high finesse optical resonators permits the selection of a limited number of well controlled degrees of freedom with which to monitor many-body systems  \cite{niranjan_measuring_2019,sawant_lasing_2017}. 

In this paper, we experimentally demonstrate the collective self-trapping of atoms in a laser-driven optical resonator. Self-trapping has been demonstrated before based on the nonlinear interaction between ultracold atoms in a bosonic Josephson junction \cite{albiez_direct_2005,abbarchi_macroscopic_2013,reinhard_self-trapping_2013}. In our system however, the dynamically coupled cavity field is the key to the effect. The laser frequency is tuned away from the mode resonance such that the light cannot penetrate into the optical resonator. However, when atoms are present, the mode frequency is pulled closer to that of the driving laser, and light is injected into the cavity. This light is far detuned from    atomic resonances, so its mechanical effect is limited to the optical dipole force. In the case of red detuning with respect to the atomic resonances, the field creates an optical dipole trap, attracting atoms towards the field maxima, \textit{i.e.} the antinodes along the cavity axis. In conservative optical dipole potentials, the trapping works for single atoms as well as for large ensembles. In contrast, in a cavity-based, dynamic, optical trap the collective effect of the atomic cloud is needed to let the necessary amount of light into the cavity. Therefore, whereas a single atom would not be trapped, a cloud with a sufficient number of atoms will be captured in the cavity mode volume under the same external conditions. This is the effect we have observed.

\section{Optical dipole trap in a cavity}
\label{sec:system}

Consider the interaction of atoms with a single, standing-wave mode of a cavity in the far-detuned regime. The cavity is externally driven by coherent laser light at a frequency $\omega$, which is close to resonance with the cavity mode at frequency $\omega_C$ and linewidth $\kappa$ (HWHM). The laser is very far detuned from the atomic resonances, $\Delta_A = \omega-\omega_A$ and $|\Delta_A|  \gg \gamma$, where $\omega_A$ is the atomic transition frequency, and the corresponding linewidth is $\gamma$.  In this limit, the atoms act as a dispersive medium on the light field. In a frame rotating at the laser frequency $\omega$, the dominant interaction is described by the Hamiltonian:
\begin{equation}
\label{eq:H_dispersive}
H/\hbar = -\Delta_C a^\dag a - i \eta (a^\dagger - a ) + U_0 \, \sum_j
|f(\mathbf{r}_j)|^2  a^\dag a
\end{equation}
where $a, a^\dag$ are the bosonic operators associated with the cavity mode, the detuning $\Delta_C=\omega-\omega_C$, and $\eta$ is the effective laser driving amplitude. The mode function $f(\mathbf{r})$ is real, and normalized to have a maximum value of 1. The atomic positions are $\mathbf r_j$, where  $j = 1\ldots N$ labels the atoms.  The last term accounts for the frequency pulling effect, namely, that the atoms give rise to a shift of the cavity mode resonance, $\omega_C \rightarrow \omega_C + N_{\rm eff} U_0$, where the effective number of atoms $N_{\rm eff} = \sum_j | f(\mathbf{r}_j)|^2 \leq N$, and the single-atom light shift $U_0= g^2 \Delta_A/(\Delta_A^2 + \gamma^2) \approx g^2/\Delta_A$ with $g$ denoting the single-photon Rabi frequency in the cavity mode. In turn, with regards to the atomic motion,  this interaction term corresponds to an optical dipole potential with spatial form determined by the mode function $| f(\mathbf{r})|^2$. The depth of the potential is proportional to the photon number. For a given distribution of the atoms, the stationary intensity in the cavity reads
\begin{equation}
 \label{eq:intensity}
 \ev{a^\dagger a} = \frac{\eta^2}{\qty(\Delta_C-N_{\rm eff}  U_0)^2+\kappa^2} \; .
 \end{equation}

The choice of the atomic detuning, $\Delta_A$, is crucial to the self-trapping effect. On the one hand, the smaller the magnitude of $|\Delta_A|$, the larger the frequency pulling effect characterized by the light shift parameter $U_0 \propto \Delta_A^{-1}$. The change of the intracavity intensity between the  atom-filled and the empty cavity cases is
\begin{equation}
\delta \langle a^\dag a\rangle = \frac{\eta^2}{\Delta_C^2 + \kappa^2} \frac{(2\Delta_C - N_{\rm eff} U_0 ) N_{\rm eff} U_0}{(\Delta_C - N_{\rm eff} U_0 )^2 + \kappa^2}\,,
\end{equation}
\textit{i.e.} the intensity becomes more sensitive to the number of atoms, $N_{\rm eff}$, for increasing $U_0$. On the other hand, the trapping time in an optical dipole trap is proportional to  $|\Delta_A|$. This can be readily seen by considering the trap depth and heating rate. The trap depth is
\begin{equation}
\label{eq:PotentialDepth}
V_{\rm dip} \approx \hbar |\Delta_A| \, \frac{g^2}{\Delta_A^2 + \gamma^2} \, a^\dag a = \hbar  |\Delta_A| \, s,
\end{equation}
where the saturation parameter, $s$, is used in order to express the field intensity and atom-field coupling in a simple form. This form of the trap depth is valid in the low saturation regime. Here the parameter $s$ corresponds to the maximum saturation of atoms reached at the antinode of the field. Although we consider the far-detuned regime, the residual spontaneous scattering of photons leads to a heating of atoms \cite{Murr2006Momentum}, described by the rate \cite{Domokos2001Semiclassical},
\begin{equation}
\label{eq:RecoilHeating}
 D = \frac{3}{10} \, \hbar \omega_{\rm rec} \gamma s,
\end{equation}
where $\omega_{\rm rec} = \hbar k^2/2m$ is the recoil frequency, $k$ and $m$ are the wavenumber and mass, respectively, and the 3/10 prefactor comes from the geometry and polarization \footnote{The recoil heating coefficients 2/5 and 1/5, this latter is along the atomic polarization,  in the two directions perpendicular to the cavity axis were averaged. Heating along the cavity axis is disregarded since it does not take the atom out of the cavity mode.}. Like the trap depth, the heating rate also scales linearly with the atomic saturation. Therefore, in the idealized case, where the heating originates solely from the weak fluorescence, the time an atom stays in an optical trap can be estimated as
\begin{equation}
 \label{eq:TrappingTime}
 \tau_{\rm trap} \approx\frac{V_{\rm dip}}{D} =  \frac{10}{3} \, \frac{ |\Delta_A|}{\gamma \, \omega_{\rm rec}}  \,,
\end{equation} 
which is independent of the light intensity. Only the detuning is a free parameter. A practical trade-off for the detuning is in the range of $\Delta_A \sim -2\pi \cdot  1$ GHz in our experimental system, resulting in $ \tau_{\rm trap} \approx50$ ms for ${}^{87}$Rb from Eq.~(\ref{eq:TrappingTime}). Since atom-atom interactions mediated by spontaneous photon rescattering and short-range collisions are negligible at the given density of atoms in the cloud, this trapping time characterizes single atoms as well as atomic ensembles. 

In general, the estimate of the trapping time in Eq.~(\ref{eq:TrappingTime}) is oversimplified: particularly when considering saturation and other heating processes. If the saturation $s$ is too low then the cavity trapping potential is not strong enough to counteract the kinetic motion at the given initial energy, $k_B T$. When increasing the saturation, the simple linear dependence of the potential in Eq.~(\ref{eq:PotentialDepth}) and the heating rate in Eq.~(\ref{eq:RecoilHeating}) expands to $s/(1+s)$. Besides the recoil  heating, there are other sources that can be technical in origin. The major source is the fluctuation of the cavity detuning $\Delta_C$,  due to the finite laser linewidth and the imperfect cavity lock system,  which is felt as intensity noise within the resonator. The fluctuating dipole trap potential leads to heating which has a rate dependent on the intensity, on the detuning and on other system parameters. Rather than analyze these processes in detail, we introduce an empirical heating rate, $D_0 + D_1 s$, which is a power expansion in terms of the saturation. Thus the trapping time can be estimated as
\begin{align}
 \label{eq:TrappingTimeRealistic}
 \tau_{\rm trap} &\approx\frac{V_{\rm dip} - k_B T}{D} \nonumber\\
 &\approx  \frac{\hbar |\Delta_A| \, \frac{s}{s+1} - k_B T}{ \frac{3}{10} \, \hbar\gamma \, \omega_{\rm rec} \, \frac{s}{s+1} + D_0 + D_1 s  }  \,.
\end{align} 
This estimate accounts for  the trapping time increasing linearly with $s$ for a range of low saturation, above some threshold level.  In the opposite limit, for large atomic saturation ($s\gg1$),  the dominant effect is the large intensity fluctuations of the dipole trap potential, described by $D_1 s$, that reduce the trapping time. 

This discussion illustrates that the analytical treatment of the intensity dependent trapping time is highly involved. It is more straightforward to  measure the trapping time as a function of the drive power.  In what follows we will experimentally identify the drive laser parameters needed for demonstrating  self-trapping.

\section{Experimental method}

\begin{figure}[thb]
 \includegraphics[width=0.66\columnwidth]{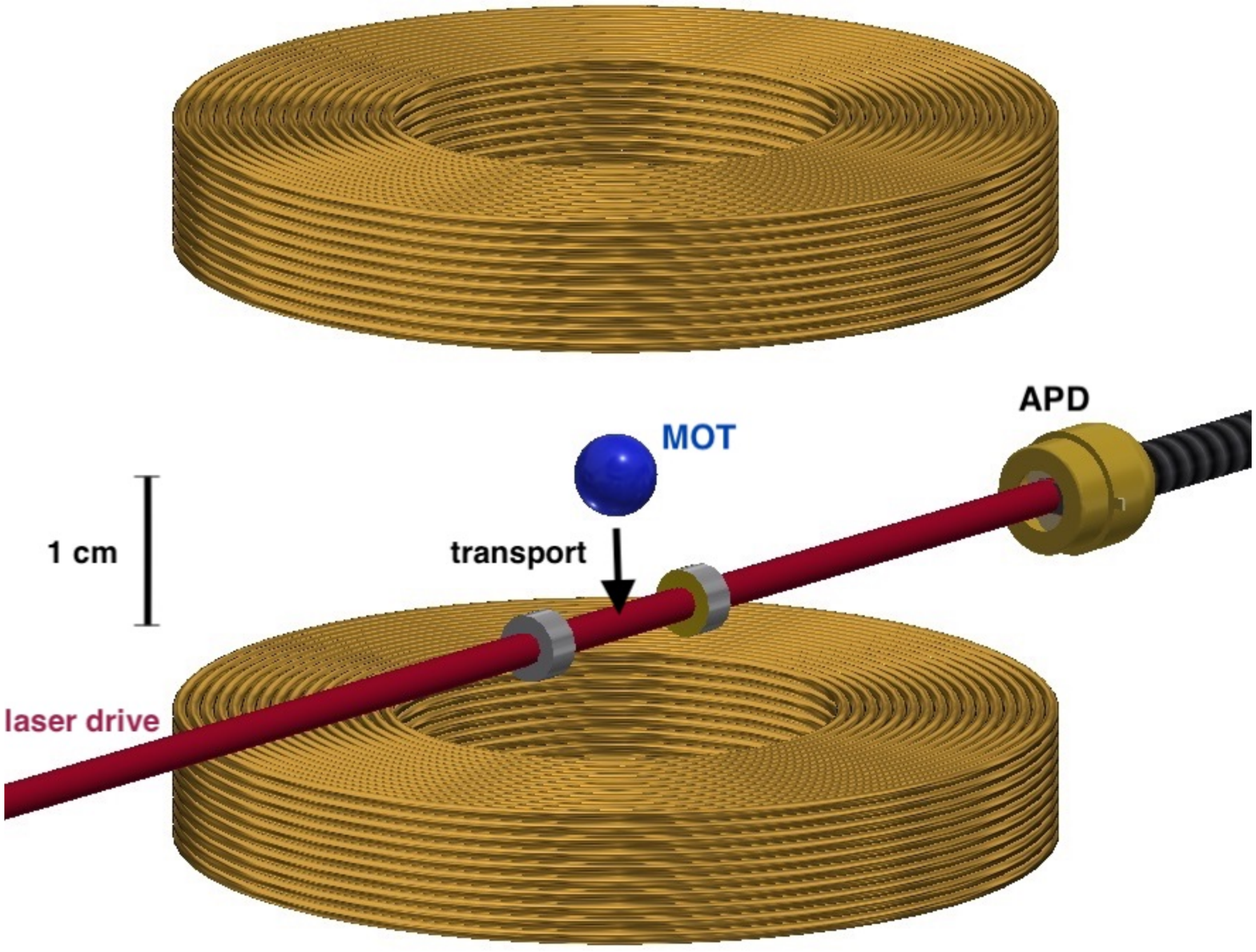}
\caption{Scheme of the experimental setup. The laser driven cavity transmission (red beam) is detected with an avalanche photo-diode. Cold atoms (blue sphere) are prepared in a magneto-optical trap and are transported by means of a magnetic trap into the cavity mode.}
\label{fig:scheme} 
\end{figure}
Our experimental setup includes a magneto-optical trap (MOT) above a high finesse optical resonator in an ultra-high vacuum chamber at a base pressure of $10^{-11}$ mbar. Rubidium atoms are captured from vapour and pre-cooled in the MOT. Then we apply polarization gradient cooling to reach typically a temperature of $T \approx 100 \mu$K. Following an optical pumping cycle, the magnetically polarized sample of cold ${}^{87}$Rb atoms in the $F=2$, $m_F=2$ hyperfine ground state is loaded into a magnetic quadrupole trap with 66 G cm${}^{-1}$ axial and 33 G cm${}^{-1}$ transverse  gradient. By varying currents in the electromagnets, the magnetic trap center can be shifted in a controlled way, allowing for vertical transport of $\sim 1$ cm of the atoms into the horizontally aligned cavity. The resonator has relatively large access from the direction transverse to the cavity axis, due to the cavity length $l=15$ mm (Fig.~\ref{fig:scheme}). The mode waist  $w=127 \mu$m is an order of magnitude smaller than the size of the atomic cloud in the direction. An effective atom-number up to the range of $N_{\rm eff} \sim 10^5$ could be achieved within the cavity mode volume.  The cavity-mode linewidth was $\kappa=  2\pi \cdot 2.77$ MHz, and the single-atom coupling constant to ${}^{87}$Rb was $g= 2\pi \cdot 0.33$ MHz on the $F=2, m_F=2 \leftrightarrow F=3, m_F=3$ hyperfine transition of the D2 line. 

In the experimental scheme, the frequency-selected fundamental Gaussian mode of the linear resonator was driven by a mode selecting laser beam through the in-coupling mirror.  The resonator was actively locked  with a variable frequency offset to a transfer cavity with sub MHz mode linewidth, which was itself locked to an atomic resonance.  The driving laser was also referenced to the same atomic resonance, thus the detuning $\Delta_C$ was variable and controlled with $\pm 200$ kHz precision. The driving laser was detuned far off the F=2 $\leftrightarrow$ F=3 atomic resonance resulting by  $\Delta_A=-2\pi \cdot 1066$ MHz. Taking into account that the other dipole allowed transitions,  F=2 $\leftrightarrow$ F=2 and  F=2 $\leftrightarrow$ F=1, also contribute to the dispersive interaction, the effective value of the light shift was calculated to be $\bar U_0 \approx 0.7 g^2/\Delta_A$ including averaging over the magnetic sublevels. The total light shift was set around $N_{\rm eff} \bar U_0 \approx -2\pi \cdot 1$ MHz, thus the cavity mode resonance was adjusted to have a detuning  in the range of $-8 < \Delta_C / 2\pi < -2$ MHz.   The intensity of the driving was varied in order to explore the full range from the weak intensity limit, where the resulting optical trapping potential is just enough to dominate gravity and the thermal motion of the atoms within the cloud, up to the high intensity limit, where the atoms were saturated.

\begin{figure}[htb]
\begin{center}
\includegraphics[width=0.89\columnwidth]{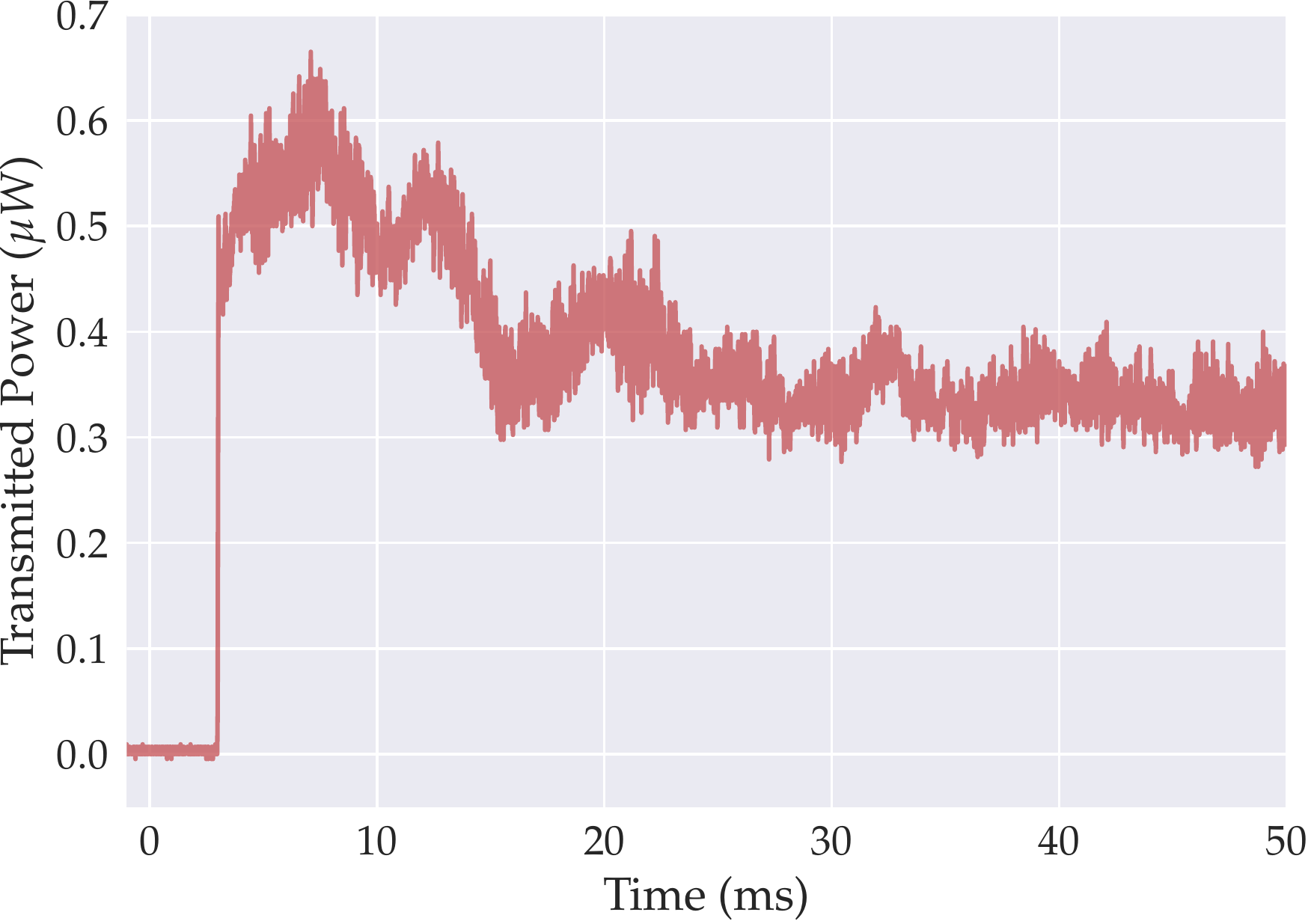}
\includegraphics[width=0.89\columnwidth]{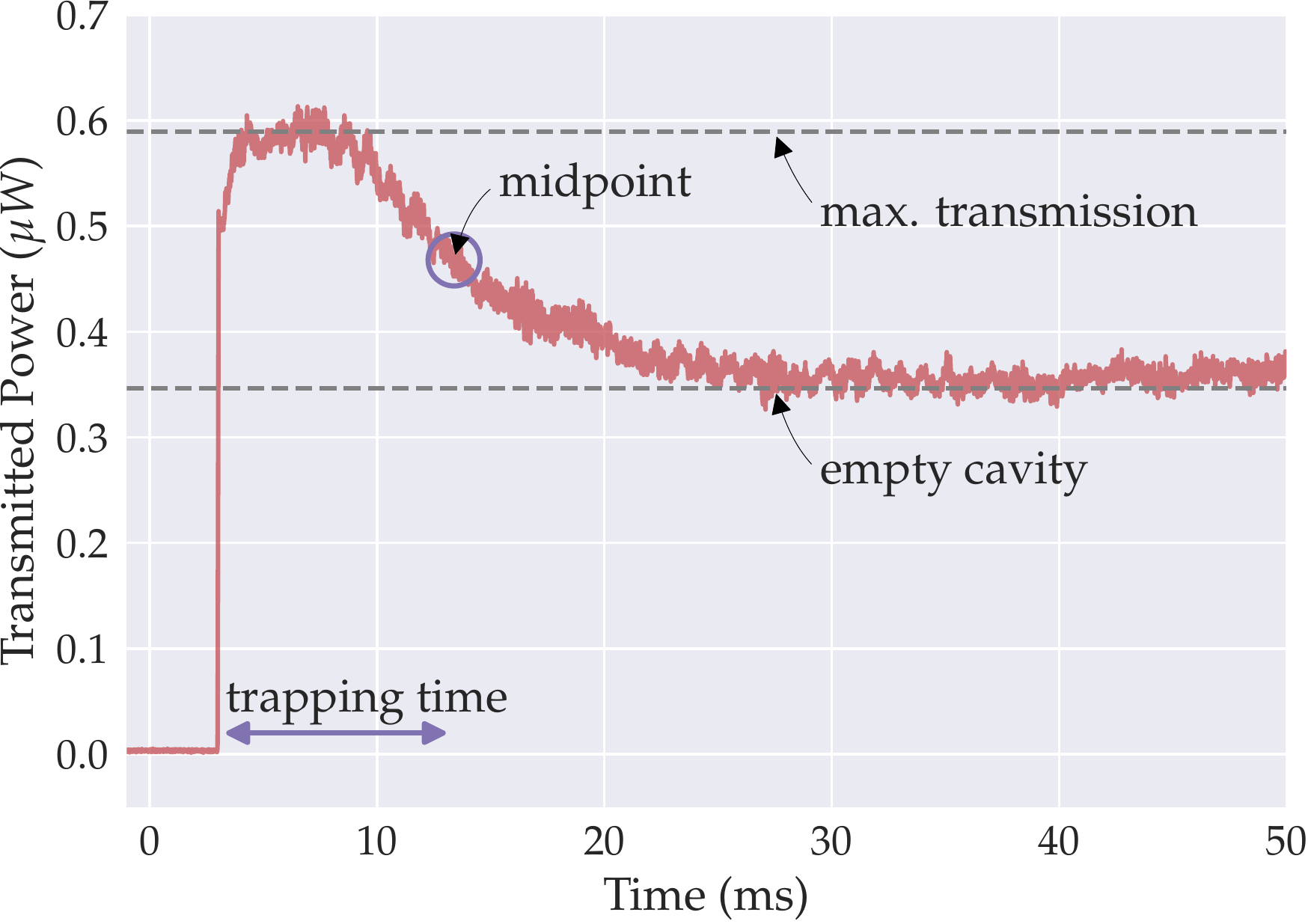}
\end{center}
\caption{Time evolution of the cavity output power with 5 $\mu$s resolution. The top panel shows a single recorded trajectory, where the non-trivial motional dynamics of the atomic cloud in the dynamically changing cavity field leads to various oscillatory features. As this depends heavily on uncontrolled initial conditions, averaging over ten such trajectories reveals a simple deterministic evolution, as shown in the bottom panel. Throughout, $\Delta_C = -2\pi \cdot 3$MHz, $N_{\rm eff} U_0 = -2\pi \cdot  1$ MHz and $\eta/\kappa = 620$.}
\label{fig:Transmission}
\end{figure}
The measurement protocol was the following. The atoms were transported from the MOT center to the cavity axis, the transport ending at $t=0$, when the magnetic trap was gently switched off over a time interval of 3 ms. Thus the thermal atom cloud was free to expand until the instant when the cavity driving was switched on at $t=3$ ms. The intracavity intensity, built up from the drive, reinstated a trap for the atoms by the optical dipole force of the cavity mode. Throughout, the transmitted intensity was detected by an avalanche photodiode (APD). Trajectories of transmitted intensity were recorded up to $t=5$ s.  A typical example of the initial evolution between  $t=0$ and $t=50$ ms is shown in Fig.~\ref{fig:Transmission} (upper panel). At $t=3$ ms,  trajectories start with an initial transient rise of the duration of $200 \mu$s  that corresponds to the opening time of the probe light's shutter. 
Then, the intensity exhibits a temporal modulation due to the displacement of the atoms induced by the gradient force of the cavity field. The corresponding signal has a randomness arising from the initial distribution of the atoms in the mode. The emergence of the trapping potential leads to a synchronized acceleration of the atoms toward the antinodes along the cavity axis. Thus the shift of the mode toward resonance with the drive is increased, which is reflected by the peak at around $t=8$ ms. This collective motion of the atom cloud is better revealed by averaging the transmission trajectories over several experimental runs, resulting in the smoother curves exhibited in Fig.~\ref{fig:Transmission} (bottom panel). One can clearly see that the atoms are held in the cavity over a time longer than 10 ms, which is in the same range as the trapping time of a conservative optical dipole trap with the same atomic detuning, $\Delta_A / 2\pi \sim -1$GHz.

\section{Demonstration of self-trapping in a cavity}
\label{sec:demonstration}

In order to study the self-trapping effect quantitatively, we need to define a characteristic trapping time. We opt for the definition that it be the interval between the point of maximum transmission and the midpoint between that and minimum transmission, corresponding to the empty cavity. An outline of the calculation can be found in the bottom panel of Figure~\ref{fig:Transmission}, for a given set of parameters. This simple  definition is chosen to facilitate the determination of short trapping times but also because the decay, as we will see later,  is not exponential. 

\begin{figure}[thb]
\centering
\includegraphics[width=0.89\columnwidth]{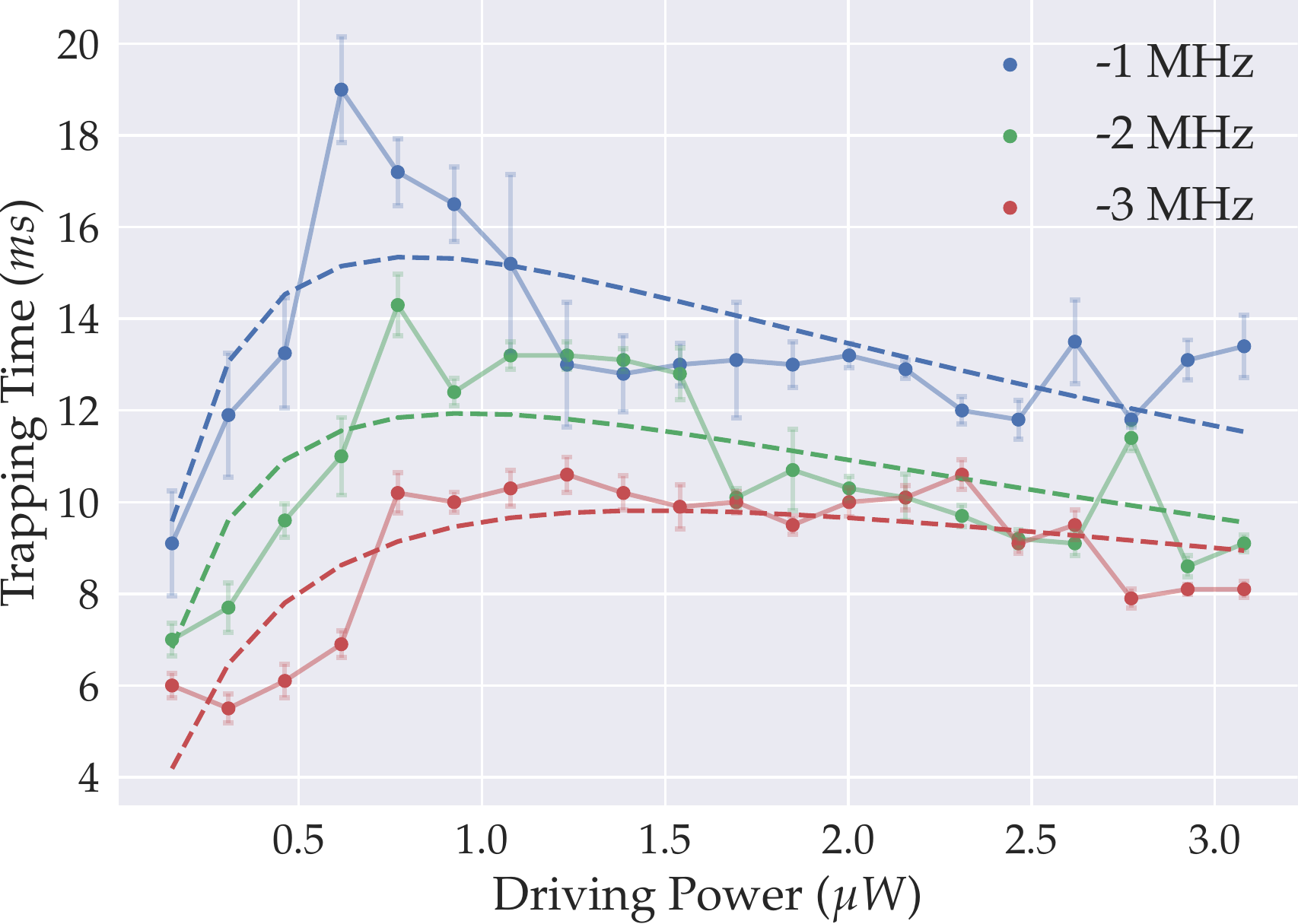}
\caption{The characteristic trapping time as a function of laser driving power  for detunings between the driving frequency and the cavity mode resonance frequency. The intensity on the horizontal axis was measured at the cavity output for an empty cavity. Connected points represent the measured trapping times extracted from the measurement data. Dashed lines represent the phenomenological model of Eq.~(\ref{eq:TrappingTimeRealistic}), with coefficients $(D_{0}=0.475, D_{1}=0.759)$ obtained from optimization for $\Delta_C=-2\pi \cdot 1$~MHz (blue dashed line). Similarly, coefficients $(0.627, 1.12)$ and $(0.884, 1.32 )$ were found for $\Delta_C=-2\pi \cdot 2$~MHz (green dashed line) and $\Delta_C=-2\pi \cdot 3$~MHz (red dashed line) respectively, where all heating rates are expressed in units of  $\tfrac{3}{10} \hbar\omega_{\rm rec} \gamma$.}
\label{fig:TrappingtimeDc} 
\end{figure}
With this definition, the effect of  atomic saturation on the trapping can be explored experimentally. 
The measured trapping times as a function of the intensity for various settings of the cavity detuning, $\Delta_C$, are presented in  Fig.~\ref{fig:TrappingtimeDc}. The overall dependence reveals three key features: (i) a minimum drive intensity to reach noticeable trapping within the cavity mode volume is required, (ii) the trapping time increases with intensity up to an optimum, from where (iii) it decreases due to atomic saturation effects. The optimum power, $P\approx 0.7 \, \mu$W, corresponds to a drive intensity for which $s\approx 0.02$. The measured data thus confirm the qualitative considerations at the end of Sect.~\ref{sec:system}. The phenomenological model presented in Eq.~(\ref{eq:TrappingTimeRealistic}) was also fitted to the measurement data, the corresponding curves are plotted with dashed lines.  The fit was restricted to the unknown heating parameters, $D_0$ and $D_1$. The model accounts nicely for the overall trend of the trapping time as a function of the input intensity.  The peak  of the measurement data near the longest trapping time,  however, seems to be missing, possibly becasue the heating associated with the dipole trap potential fluctuations is more sensitive to the intensity via the atomic saturation effect than the simple linear law used in the model. Note that the frequency fluctuations are transformed to intensity via the Lorentzian resonance filter. More detailed theoretical modelling of this effect is beyond the scope of this paper and we settle for determining the function experimentally.

\begin{figure}[thb]
\centering
\includegraphics[width=0.81\columnwidth]{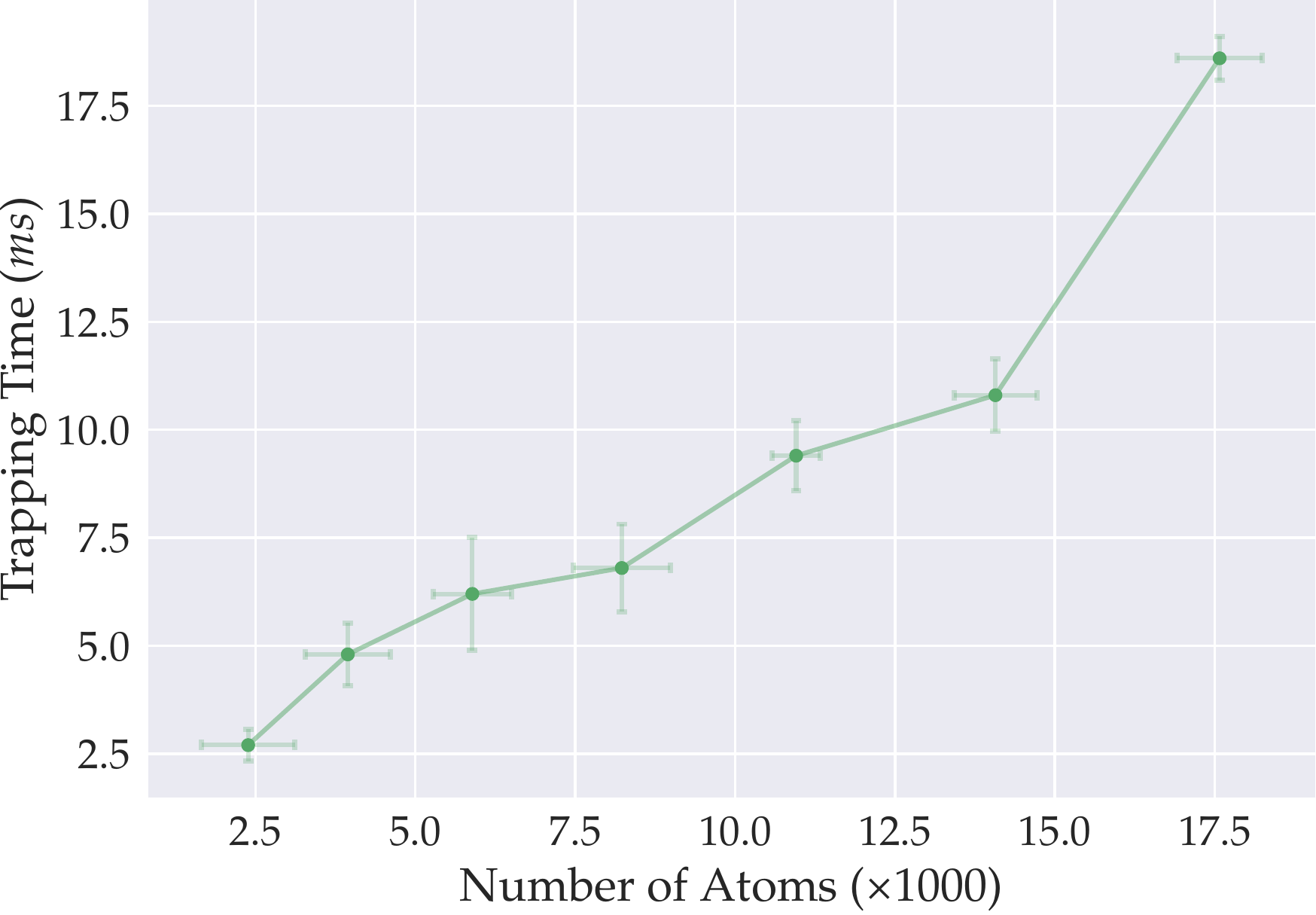}
\caption{A demonstration of trapping dilation as a function of atom number for fixed detuning, $\Delta_C=-2\pi \cdot 2$~MHz, and driving amplitude, $\eta/\kappa = 290$, with the third-smallest drive power, $P=0.46 \mu$~W, of Fig.~\ref{fig:TrappingtimeDc}. }
\label{fig:CollectiveSelftrapping} 
\end{figure}

As can be seen in Figure~\ref{fig:TrappingtimeDc}, the trapping time exhibits a significant dependence on both the intensity and the cavity detuning in the low saturation regime. Fixing the parameters of the driving laser, \textit{i.e.} its frequency and intensity, we measure the trapping time as a function of the atom number. To this end, the number of atoms transported into the cavity mode can be varied by using different initial loading of the MOT. The actual atom number in the cavity can then be calibrated by comparing the transmission at the initial time, $t=3$ms, and at late times when the cavity is empty, {\it e.g.} at $t> 4$s. The result is shown in Fig.~\ref{fig:CollectiveSelftrapping}.
The increase of the trapping time for increasing atom number clearly demonstrates the collective nature of the self-trapping effect. In a conservative potential, the trapping time is independent of the number of atoms to leading order. The weak effect of  atom-atom collisions is expected to reduce the trapping time. Here we observe the opposite. The enhanced trapping can be attributed to the increased intensity which results from the atoms pulling the cavity mode resonance closer to the drive frequency. As can be seen in Fig.~\ref{fig:TrappingtimeDc}, there is an optimum drive power for maximizing the trapping time, and this optimum is reached by the frequency pulling effect.  Moreover, the amount of frequency pulling is proportional to the number of atoms, hence the intensity as well as the trapping time increases with the atom number. This is clearly a collective effect. A single atom would not be trapped under the same driving conditions. The atoms collectively help create the trap, which in turn holds them captive for $\sim 20$ ms, in line with our expectations for the given atom-light detuning of $\sim - 1 $GHz.

\section{Collapse of the trap: non-exponential decay}

The counterpart of the collectively self-created trap is that the slow evaporation of atoms from the trap sets off its self-destruction. As the number of atoms decreases, the trapping conditions degrade \textit{i.e.} the trap depth is reduced.
This in turn facilitates the escape of atoms from the dipole trap, triggering a positive feedback, which leads ultimately to an accelerated, non-exponential decay of the atomic cloud in the cavity.

The resulting non-exponential decay can be described phenomenologically by a simple model inspired by the Metropolis–Hastings algorithm. The evolution of the atom number, $N$, in the trap is followed in discrete steps. The potential felt by the atoms is of depth
\begin{equation}
 V(N)=U_0 \ev{a^\dagger a}(N)\, ,
\end{equation}
where the atom-number dependent intensity is given in Eq.~(\ref{eq:intensity}). Further, we introduce an inverse temperature, $\beta$, for the atomic cloud. Then the probability for a single atom to leave the trap in a single time step is
\begin{equation}
\label{eq:escapeRate}
 p(N)=\frac{N}{\tau} \,  e^{-\beta V(N)} =\frac{N}{\tau} \,   e^{-\frac{\mathcal A}{\qty(\tilde\Delta_C-N \tilde U_0)^2+1}}\,,
\end{equation}
where the quantities with tilde are measured in units of $\kappa$ and $1/\tau$ is a decay constant setting the time scale of the decay. The dimensionless parameter, $\mathcal A$, has the character of an intensive thermodynamic property, and incorporates several parameters of the system, \textit{e.g.} drive amplitude, temperature, \textit{etc.} Without the exponential factor ($\mathcal A=0$), an exponential decay of the atom number from the trap would be expected, but that factor modifies this behaviour.

\begin{figure}[thb]
\centering
\includegraphics[width=0.81\columnwidth]{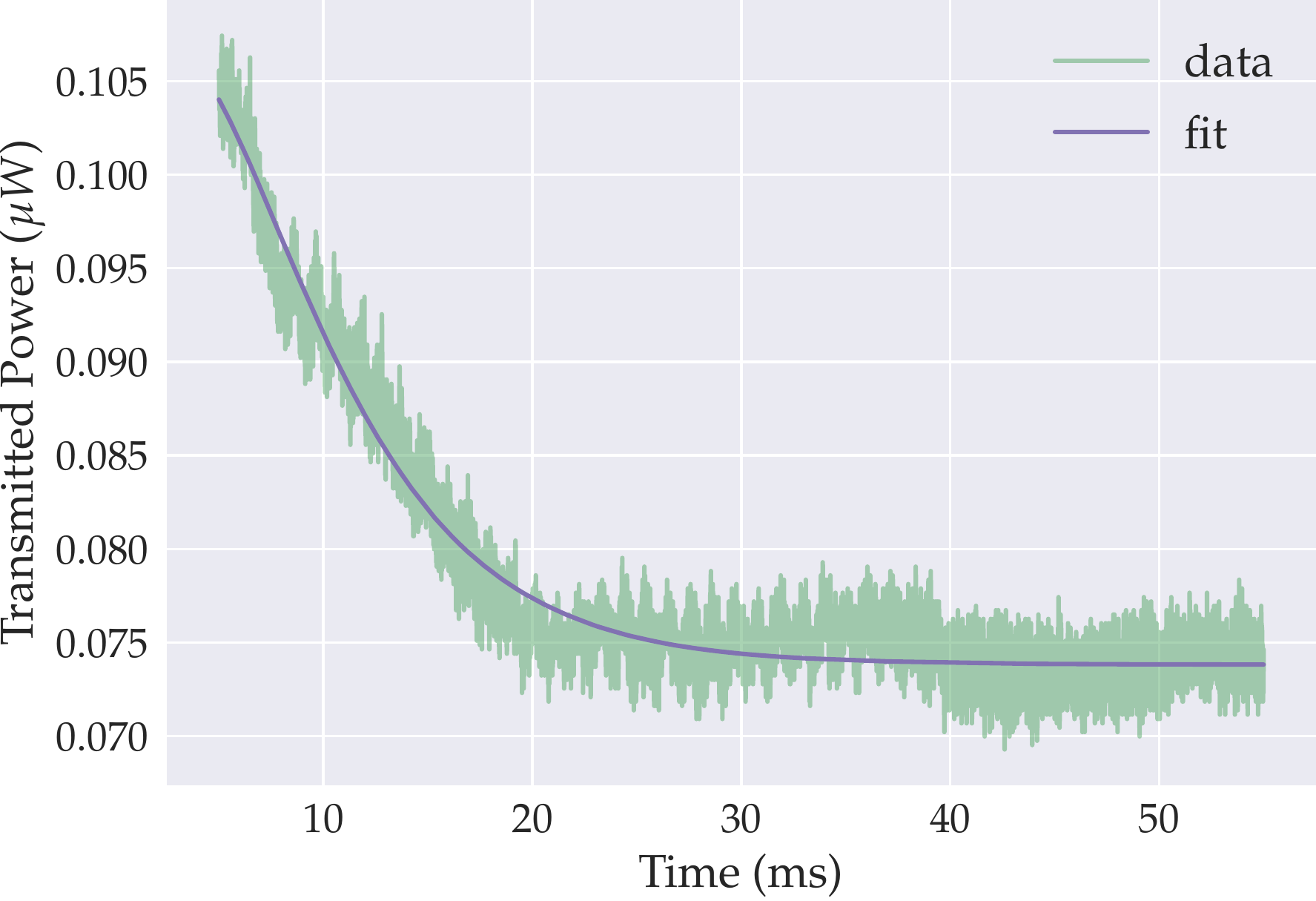}
\caption{The temporal self-destruction of the collective trap, as seen from the cavity transmission. The fitting relies on the simulated atom number for atoms undergoing non-exponential decay.  For the parameter $\mathcal A$, a significant non-vanishing value is extracted from the fit, confirming the non-exponential nature of the decay of the atom number in the trap.}
\label{fig:collapse} 
\end{figure}

We generated Monte Carlo trajectories consisting of random sequences of escape times. We found that averaging on the order of 10 such trajectories yielded smooth curves, for sufficiently high initial atom numbers, that could be fit to the observed data. Such a fit is presented in Fig.~\ref{fig:collapse}.

All the parameters of Eq.~(\ref{eq:escapeRate}) have been optimised for the best fit. We obtained the detuning, $\Delta_C = -2\pi \cdot 1.87$ MHz, and $N(0) U_0= -2\pi \cdot 1$ MHz which are in good agreement with the measured experimental values, i.e., we have the same atom number and $U_0$ as in the  case of Fig.~\ref{fig:Transmission}. The key result is that the parameter  $\mathcal A = 2.775$ is significant, thereby the fit demonstrates strongly non-exponential character of the decay.

\section{Conclusions}

The  experiment demonstrates optical dipole trapping of a cloud of cold atoms coupling them to a single mode of a high-finesse optical cavity. The detuning between the atoms and the mode is adjusted to reach a considerable capture time in the dipole trap. At the same time,  the atoms still have a considerable back-action on the trapping field mode, which  serves for real-time monitoring the atoms in the trap.  The trapping manifests the consequences of frequency pulling: a well known collective effect. We explored a parameter regime where frequency pulling of the mode initiates trapping and demonstrated that, to this end, a sufficient number of atoms is required. On the other hand, by showing that the collapse of the trapping follows a non-exponential law, due to the dynamic nature of the light field in the cavity, we demonstrated both self-trapping and self-destruction, bringing our investigation full circle.

\section*{Acknowledgements}

This work was supported by the National Research, Development and Innovation Office of Hungary (NKFIH) within the Quantum Technology National Excellence Program  (Project No. 2017-1.2.1-NKP-2017-00001) and the Quantum Information National Laboratory of Hungary. D. Nagy and A. Vukics were supported  by the J\'anos Bolyai Fellowship of the Hungarian Academy of Sciences.

\bibliography{self-trapping,nagyd}

\end{document}